\documentclass[aps,prd,groupedaddress,nofootinbib]{revtex4}

\usepackage{amsmath,amssymb,graphicx,color,epsf}
\usepackage[export]{adjustbox}
\usepackage{comment}
\usepackage{natbib}
\usepackage[scanall]{psfrag}
\usepackage{tikz}

\begin{document}

\title{Primordial perturbations in Type III hilltop inflation models}

\author{Chia-Min Lin$^{1}$}
\email{cmlin@ncut.edu.tw}
\author{Harish Dhananjay Nalla$^{2}$}
\email{810914305@gms.ndhu.edu.tw}
\author{Chen-Pin Yeh$^{2}$}
\email{chenpinyeh@gms.ndhu.edu.tw}
\author{Da-Shin Lee$^{2}$}
\email{dslee@gms.ndhu.edu.tw}

\affiliation{$^{1}$Fundamental General Education Center, National Chin-Yi University of Technology, Taichung 41170, Taiwan, R.O.C.}
\affiliation{$^{2}$Department of Physics, National Dong Hwa University, Hualien 97401, Taiwan, R.O.C.}



\begin{abstract}
We analytically compute the power spectrum of primordial curvature perturbations in Type III hilltop inflation models under the slow-roll approximation. The model parameters are constrained using current Cosmic Microwave Background (CMB) data. The curvature perturbations that exit the horizon at small scales show sufficiently large amplitudes to produce primordial black holes (PBHs). We then consider the quantum one-loop corrections in these models from both the self-interaction of the inflaton and its interaction with the waterfall field. We show the loop corrections in both cases for 60 e-folds of inflation are negligible, ensuring the tree-level results are reliable within the chosen parameter regime. 
\end{abstract}
\maketitle
\large
\baselineskip 18pt
\section{Introduction}
\label{sec1}
Cosmic inflation \cite{liddle2000cosmological,starobinsky1980new,guth1981inflationary,linde1982new} is considered the standard scenario for describing the evolution of the early universe. This simple idea solves almost all the problems of the conventional hot big bang model, such as the horizon problem, the flatness problem, the unwanted relics problem, etc. Furthermore, it amplifies quantum fluctuations to cosmological scales, providing initial conditions for structure formation and cosmic microwave background (CMB) temperature fluctuations. Observations of CMB and large-scale structures are crucial in testing inflationary models. 

Inflationary models split into large-field and small-field models based on whether the associated field value is large or small compared to the reduced Planck mass $M_P \equiv 1/\sqrt{8\pi G}$ during inflation. Large-field models predict significant magnitudes of primordial gravitational waves that are potentially detectable by CMB observations. However, despite increasing precision, no primordial gravitational waves have yet been discovered. Therefore, large-field models may not be as favored. On the other hand, small-field models are appealing because they allow for the power series expansion of the potential. In this work, we consider Type III hilltop inflation models introduced in \cite{kohri2007more} (referred to as model 3 therein). These small-field models typically produce a red spectrum and satisfy CMB constraints. Interestingly, the perturbations that cross the horizon at the end of the inflation (small scales) generally have a larger amplitude than those corresponding to the CMB scales. Those small-scale perturbations are relevant for producing PBHs and inducing gravitational waves \cite{kohri2007more, alabidi2012observable, alabidi2009generating}. 

Recently, a debate has emerged on whether the formation of PBHs during single-field ultra-slow-roll (USR) inflation could induce significant loop corrections on the large-scale perturbations probed by CMB \cite{kristiano2024constraining, riotto2023primordial, riotto2023primordialblackholeformation, choudhury2023no, choudhury2023pbh, choudhury2023quantum, firouzjahi2023one, firouzjahi2024primordial, franciolini2023one, fumagalli2023absence, syu2020quantum, cheng2022power, cheng2024primordial, stewart1997flattening, Stewart:1997wg}. One of the authors and his collaborators have explored loop effects in single-field inflation with a transient USR phase. This phase generates significant curvature perturbations at small scales that lead to the formation of PBHs \cite{syu2020quantum, cheng2022power, cheng2024primordial}. Their work gave conditions on model parameters to ensure that the curvature perturbations satisfy CMB constraints and are substantial enough to contribute to PBH formation.

In this work, we continue this line of investigation, focusing on Type III hilltop inflation models where the waterfall field acts as an assisting spectator scalar field. We analytically obtain the power spectrum under the slow-roll approximation at the tree level following \cite{kohri2007more} and impose CMB constraints to obtain conditions on model parameters. We analyze the spectrum and identify models that can produce PBHs. We then consider quantum loop effects in these models from both the self-interaction of the inflaton field and its interaction with the waterfall field. We estimate loop effects by calculating the corrections to slow-roll parameters following \cite{cheng2024primordial, boyanovsky2006quantum, sloth2006one, sloth2007one}. This method differs from the effective potential approach in flat spacetime \cite{liddle2000cosmological, stewart1997flattening, Stewart:1997wg}, as the inclusion of the Hubble parameter as an additional energy scale results in significantly different quantum loop corrections \cite{boyanovsky2006quantum, syu2020quantum}. We restrict our study to the slow-roll approximation without involving other mechanisms that drive the universe away from this regime.

\section{Hybrid inflation and Type III hilltop inflation}
\label{hybrid}
In this section, we review hybrid and Type III hilltop inflation models and impose CMB constraints to determine conditions on model parameters of Type III hilltop models. We set the reduced Planck mass $M_P =1$ throughout this work. During inflation, the inflaton field in hybrid inflation \cite{linde1994hybrid, copeland1994false}  has an effective potential
\begin{equation}
V=V_0+\frac{1}{2}m^2 \phi^2,
\label{hy}
\end{equation}
namely, a mass term of $\phi$ plus a constant potential term. One of the main purposes of hybrid inflation is to build a small-field model, which requires $V \simeq V_0$ to satisfy the following two slow-roll conditions. 
The first slow-roll condition is
\begin{equation}\label{epsilon}
\epsilon_{1v} \equiv \frac{1}{2} \Bigg( \frac{V^{\prime}(\phi)}{V}\Bigg)^2  \ll 1,
\end{equation}
and the second slow-roll condition is
\begin{equation} \label{sr2}
\epsilon_{2v} \equiv \Bigg(\frac{V^{\prime\prime}(\phi)}{V}\Bigg)   \ll1.
\end{equation}

The addition of $V_0$ to the potential $V$ can be effectively achieved by introducing a waterfall field $\psi$ with a Higgs-like symmetry-breaking potential
\begin{equation}
V(\phi,\psi)=\frac{1}{2}m^2 \phi^2 +g^2\phi^2\psi^2 +\kappa^2(\psi^2-\Lambda^2)^2.
\end{equation}
The effective mass of the waterfall field can be read off from the quadratic term as
\begin{equation} \label{effectivemass}
M_{\psi}^2(\phi)=\partial_{\psi}^2 V=2g^2\phi^2-4\kappa^2 \Lambda^2.
\end{equation}
The inflaton field starts from a non-zero initial value and moves towards the minimum of its effective potential. During inflation when $ \phi > \phi_e$, the inflaton field gives a sizeable positive mass to $\psi$ 
where
\begin{equation}
\phi_e=\frac{\sqrt{2}\Lambda\kappa}{g}
\label{g}
\end{equation}
is determined by $\partial_{\psi}^2V\big|_{\psi=0~\phi=\phi_{e}}=0$.
Therefore, the waterfall field is trapped at false vacuum $\psi=0$, and the inflaton potential is of the form given by Eq.~(\ref{hy}) with $V_0=\kappa^2 \Lambda^4$. Inflation ends at $\phi_e$, after which the waterfall field becomes tachyonic with negative mass squared. To avoid a second stage of inflation driven by the symmetry-breaking potential of the waterfall field, we require the curvature at the top of this potential to be large enough such that at $\phi=0$
\begin{equation} 
\left|\frac{\partial_{\psi}^2V}{V}\right|=\frac{4}{\Lambda^2} \gg 1,
\label{L}
\end{equation}
giving the constraint $\Lambda \ll 2$. If there is a second stage of inflation during the waterfall phase, the fluctuations of the waterfall field may contribute to the primordial perturbations at tree-level \cite{clesse2015massive, kawasaki2016can, tada2023primordial}. A numerical study of the waterfall phase is given in \cite{clesse2011hybrid} (see also \cite{Kodama:2011vs}). \newline
\begin{figure}[t]
  \centering
\includegraphics[width=0.6\textwidth]{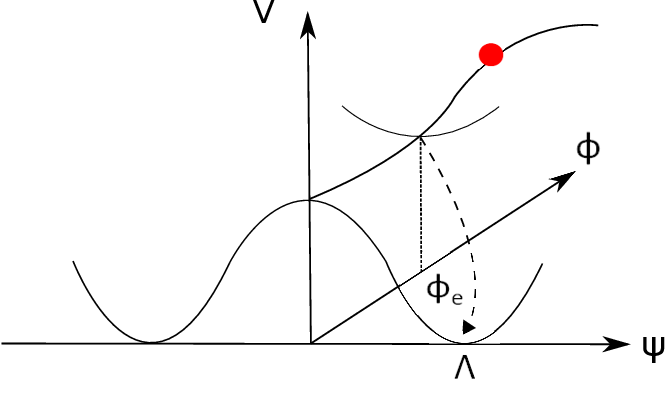}
  \caption{The typical potential for type III hilltop inflation.}
  \label{fig_pot}
\end{figure}

\begin{figure}[t]
  \centering
\includegraphics[width=0.6\textwidth]{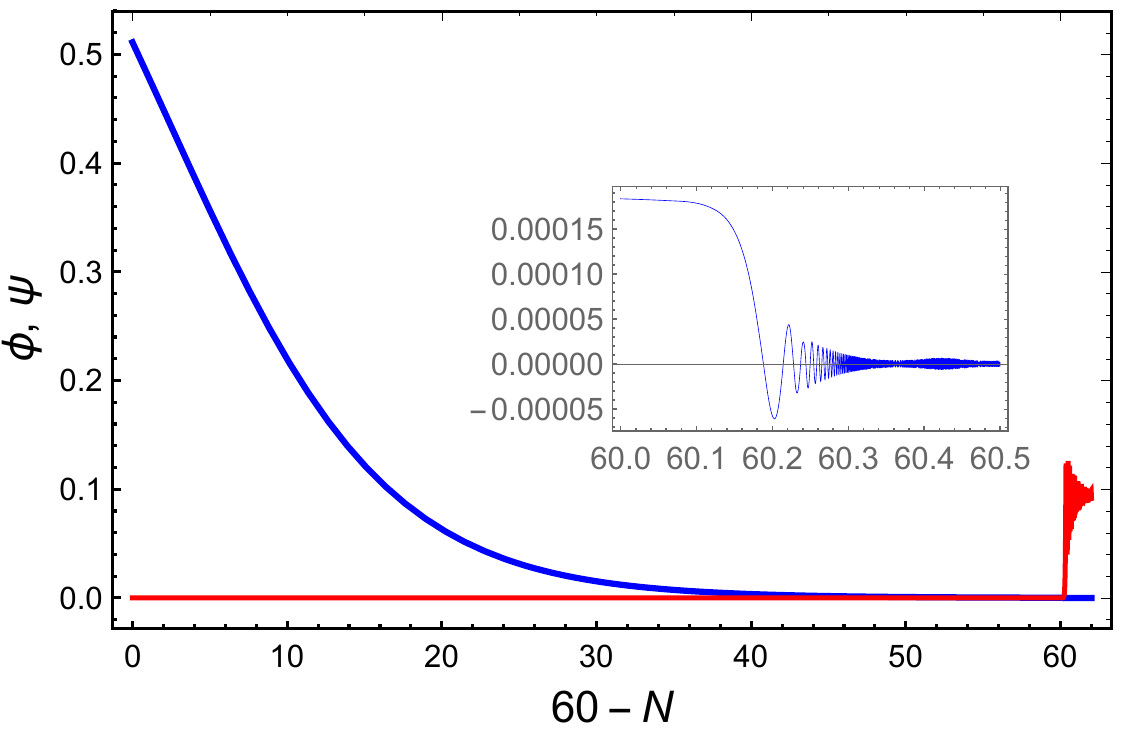}
  \caption{The time evolution of inflation field $\phi$ (blue), and the waterfall field  $\psi$ (red) as a function of e-fold number $N$ obtained by solving the two-field background equations \cite{gordon2000adiabatic} for the model $p=3$ with $\eta_{0}=0.14$ and $V_{0}=2.78463 \times 10^{-10}$.}
  \label{figa}
\end{figure}

The Hybrid inflation model has already been ruled out by observation \cite{akrami2020planck} because its potential is concave upwards, which produces a blue spectrum, whereas observations indicate a red spectrum. One can introduce an additional term that makes the potential concave downward, thus converting it into hilltop inflation. Here, we consider a class of hilltop inflation models with the potential of the inflaton field given by \cite{kohri2007more}: 
\begin{equation}
V_\phi(\phi)=V_0 \left( 1+\frac{1}{2}\eta_0 \phi^2 \right)-\lambda \phi^p+\cdots,
\label{v}
\end{equation} 
where $\lambda>0$ and by Type III, we mean $\eta_0>0$ and $p>2$. These small-field models are attractive because they can be realized in the framework of supersymmetric quantum field theories \cite{lin2024hilltop, kohri2014hilltop, lin2009reducing}. These models can be tested using gravitational waves if cosmic strings form due to the symmetry-breaking potential of the waterfall field after inflation \cite{lin2021testing}. The inflaton field may play the role of the Affleck-Dine (AD) field and produce baryon asymmetry after inflation \cite{lin2020inflaton}. The supersymmetric model could evade thermal and non-thermal gravitino problems and generate gravitino dark matter \cite{kohri2010hilltop}. Therefore, the ramifications are very rich and deeply connected to particle physics. The typical profile of Type III hilltop potential is given by 
\begin{equation}
V(\phi,\psi)=V_{\phi}(\phi)+g^2\phi^2\psi^2 +\kappa^2(\psi^2-\Lambda^2)^2-\kappa^2 \Lambda^4,
\label{hy2}
\end{equation}
as illustrated in Fig.~\ref{fig_pot}. Where, $V_0=\kappa^2 \Lambda^4$ is absorbed into $V_{\phi}(\phi)$ as given in (\ref{v}). We consider the scenario where the inflaton field gives a large mass squared to the waterfall field $\psi$, causing it to remain in the false vacuum state $\psi=0$. In Fig. \ref{figa}, we illustrate the evolution of fields during inflation by solving the full two-field background equations \cite{gordon2000adiabatic}. The figure shows that the waterfall field $\psi$ remains at $\psi=0$ throughout inflation. At the end of inflation, as the inflaton field reaches $\phi_{e}$, the waterfall field shifts to its true vacuum state, causing inflation to end due to tachyonic instability. During inflation when $\phi>\phi_e$, the background field dynamics is effectively determined by the inflaton field and the Friedmann equation as
\begin{align} \label{friedmann1}
    \ddot \phi(t) + 3H(t)\dot \phi(t) +\frac{\partial V_{\phi} (\phi)}{\partial \phi} =0,&\\
 H^2(t) = \frac{1}{3 } \left[ \frac{\dot \phi(t)^2}{2}  + V_{\phi} (\phi) \right].&
\end{align}
 Suppose the field shifts from an initial value $\phi_{i}$ to a final value of $\phi_{e}$ during some time interval with $ \phi_e < \phi < \phi_i$ and $0 < V(\phi_{e}) < V(\phi_{i})$, then the scale factor will increase during this period by a factor
\begin{equation}
   N  = \displaystyle\int_{\phi}^{\phi_{e}} \frac{H \, d\phi}{\dot \phi} \simeq -\displaystyle\int_{\phi}^{\phi_{e}} \frac{V_{\phi}(\phi)}{ V_{\phi}'(\phi)} \, d\phi. \label{e-fold}
\end{equation} The second equality is valid under slow-roll approximation characterized by the conditions in equation ($\ref{sr2}) ~\text{and}~ (\ref{epsilon}$). At the beginning of inflation, we set $N=N_T$, where $N_T$ is the total number of e-folds, and at the end of inflation, $N=0$.  Equation (\ref{e-fold}) can be analytically solved using $V \simeq V_0$ to obtain
\begin{equation}\label{phi_N}
\phi^{p-2}=\frac{V_0 \eta_{0} e^{N (p-2)\eta_{0}}}{V_0\eta_{0} \phi_e^{2-p}+\lambda p \left(e^{N (p-2)\eta_{0}}-1\right)}\, .
\end{equation}  The power spectrum evaluated at the time of horizon crossing under this approximation is given by:
\begin{equation}\label{P_R}
    P_{\mathcal{R}}(k)  \simeq \frac{H^2}{8\pi^2 \epsilon_{1v}}.
\end{equation}Using the background equation $3H^2 \simeq V_{0}$, we have
\begin{equation}
P_{\mathcal{R}}=\frac{1}{12\pi^2} V_0^{\frac{p-4}{p-2}} e^{-2\eta_0 N}\frac{\left[ p\lambda \left( e^{(p-2)\eta_0 N}-1 \right)+\eta_0 x \right]^{\frac{2p-2}{p-2}}}{\eta_0^{\frac{2p-2}{p-2}}(\eta_0 x-p\lambda)^2},
\label{p}
\end{equation}
with
\begin{equation}
x \equiv V_0 \phi_e^{2-p}
\label{xe}
\end{equation}
and its spectral index
\begin{equation}
n_s\equiv\frac{d \ln P_{\mathcal{R}}}
{d \ln k}\Big\vert_{k=aH} +1=1+2\eta_0 \left[ 1-\frac{\lambda p (p-1)e^{(p-2)\eta_0 N}}{\eta_0 x +p\lambda (e^{(p-2)\eta_0 N}-1)} \right].
\end{equation}
The running spectral index is
\begin{equation}
\alpha\equiv \frac{
d n_s}{
d \ln k}=2 \eta_0^2 \lambda p(p-1)(p-2)\frac{e^{(p-2)\eta_0 N}(\eta_0 x-p\lambda)}{\left[ \eta_0 x +p\lambda (e^{(p-2)\eta_0 N}-1) \right]^2}.
\label{a}
\end{equation}
Planck observations constrain the upper bound of the running spectral index $\alpha<0.02$ \cite{akrami2020planck}. This value is within the $95\%$ confidence level of TT, TE, EE+lowE+lensing constraints, including running of the running index. Furthermore, imposing the CMB normalization
\begin{equation}
P^{1/2}_{\mathcal{R}} = 5 \times 10^{-5},
\end{equation}
and the spectral index
\begin{equation}
n_s=0.96,
\end{equation}
at $N=N_{\rm cmb}$, which is set to be $N_{\rm cmb}=60$, one can constrain the parameters in the hilltop model
\begin{equation}
x=p\lambda \frac{((p-2)\eta_0-0.02)e^{(p-2)N_{\rm cmb}\eta_0}+(\eta_0+0.02)}{\eta_0(\eta_0+0.02)},
\label{x}
\end{equation}
and
\begin{equation}
2.5 \times 10^{-9}=\frac{1}{12\pi^2}V_0^{\frac{p-4}{p-2}}(\lambda p)^{\frac{2}{p-2}}\left( \frac{p-1}{\eta_0+0.02} \right)^{\frac{2p-2}{p-2}}\left( \frac{\eta_0+0.02}{\eta_0(p-2)-0.02} \right)^2.
\label{lambda}
\end{equation}
For a given $p$ in the model and a fixed value of $N=N_{cmb}$, the two equations constrain the parameters $V_{0}, \eta_0,$ and $\lambda$ from the model, along with $\phi_e$ from inflation. These constraints leave two free parameters, and we choose $\eta_0$ and $V_0$. The scalar-to-tensor ratio $r$ for small field inflation requires $r<0.01$\cite{PhysRevD.110.043521}, giving a constraint $V_{0}^{1/4} < 4.2 \times 10^{-3} M_{p}$. We set $V_{0}=\left(4.085 \times 10^{-3}\right)^4M_{p}$ in this work. Introducing the degree of freedom of the waterfall field brings three new parameters, namely $g$, $\kappa$, and $\Lambda$. However, $\kappa^2 \Lambda^4$ is the constant potential term $V_{0}$ and Eq.~(\ref{g}) is used to determine $g \Lambda$ with $\Lambda\ll2$ coming from Eq.~(\ref{L}). The coupling constant $g^2<1$ for the vacuum-dominated regime \cite{copeland1994false} and to justify perturbation methods.

We will use these model parameters to study the curvature perturbations during inflation and Eqs. (\ref{p}), (\ref{x}) and (\ref{lambda}) with the constraint in Eq. (\ref{a}) will be the key equations for our analysis in the next section.

\section{Tree-level curvature perturbations} \label{sec3}

\begin{figure}[ht]
    \centering
    \includegraphics[width=0.495\textwidth]{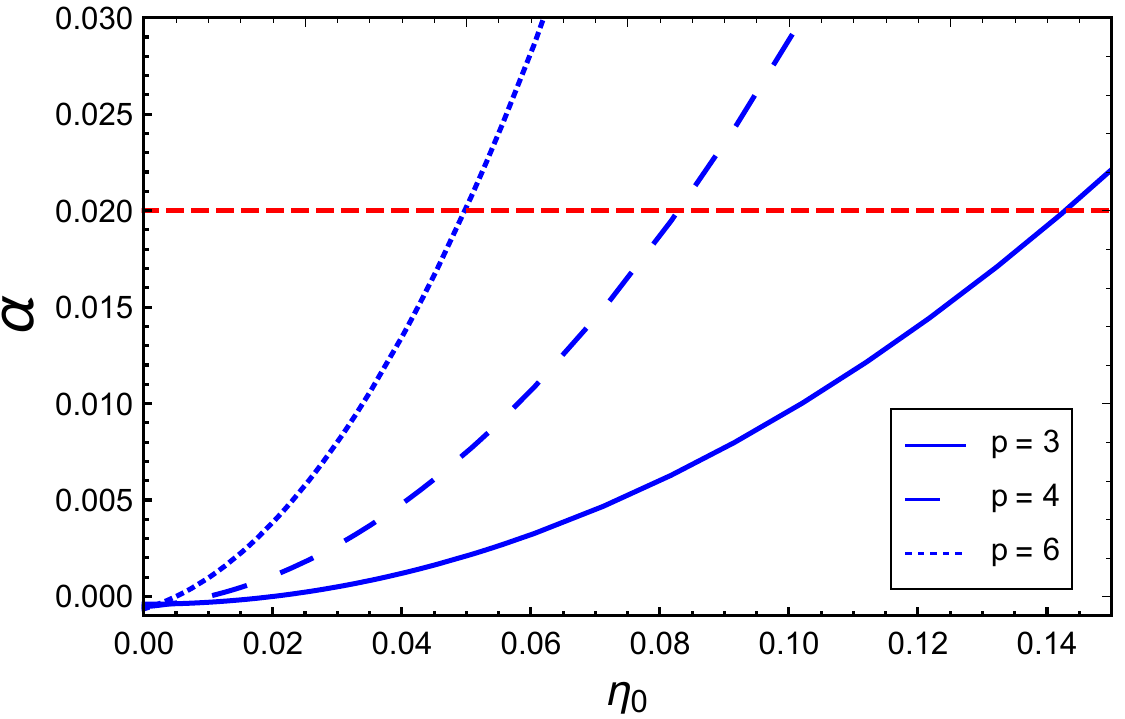} \hfill
    \includegraphics[width=0.495\textwidth]{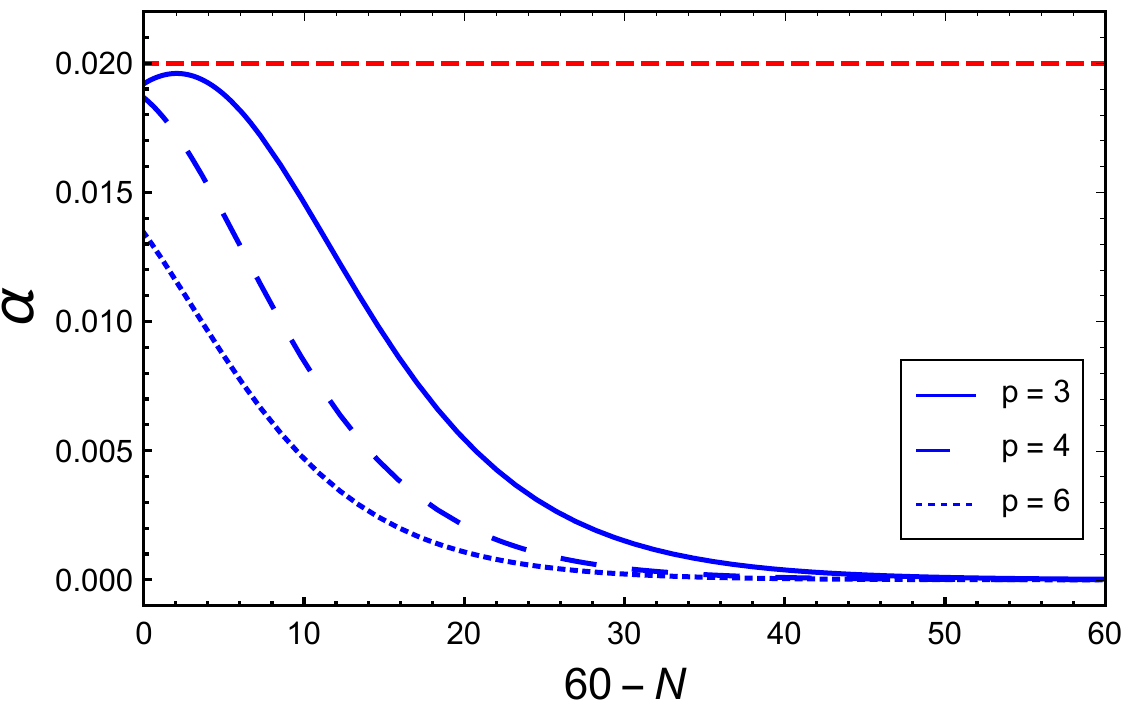}
    \caption{(a) {Plot of $\alpha$ at $N=N_{cmb}=60$ for $p=3,4$ and $6$, respectively.} The red dashed line indicates the upper bound $\alpha=0.02$. (b) Plot of $\alpha$ as a function of e-fold number $N$ where $\eta_{0}=0.14,0.08$ and $0.05$ corresponding to $p=3,4$ and $6$, respectively.}
    \label{fig0}
\end{figure}

\begin{figure}[ht]
    \centering
    \includegraphics[width=0.495\textwidth]{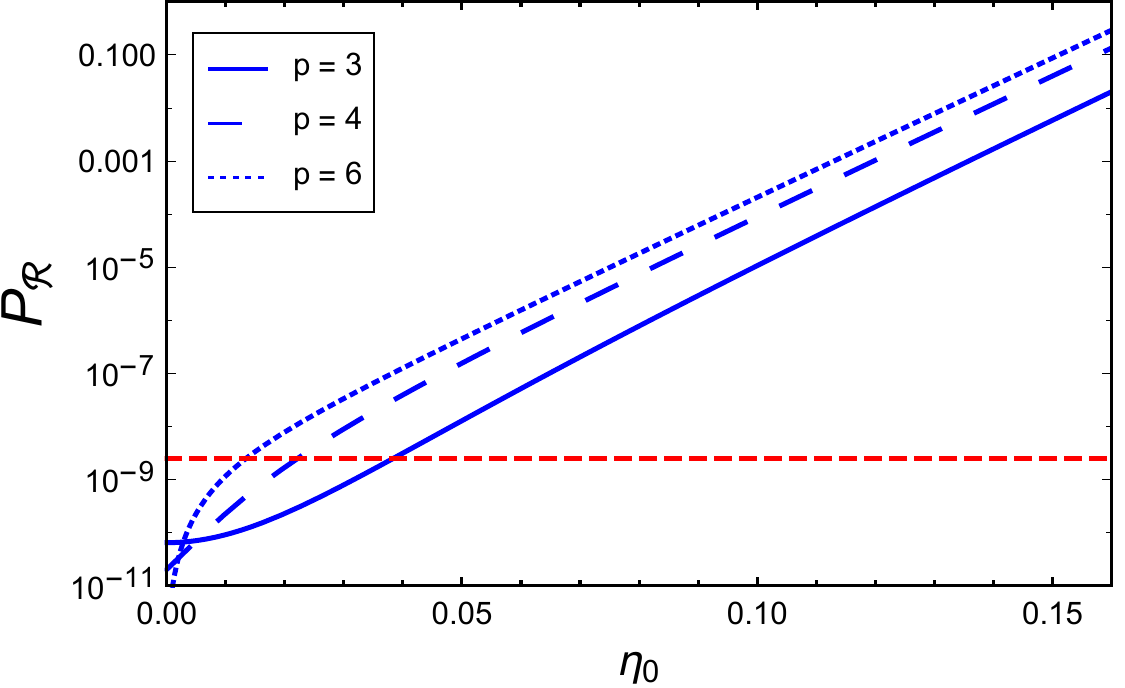} \hfill
    \includegraphics[width=0.495\textwidth]{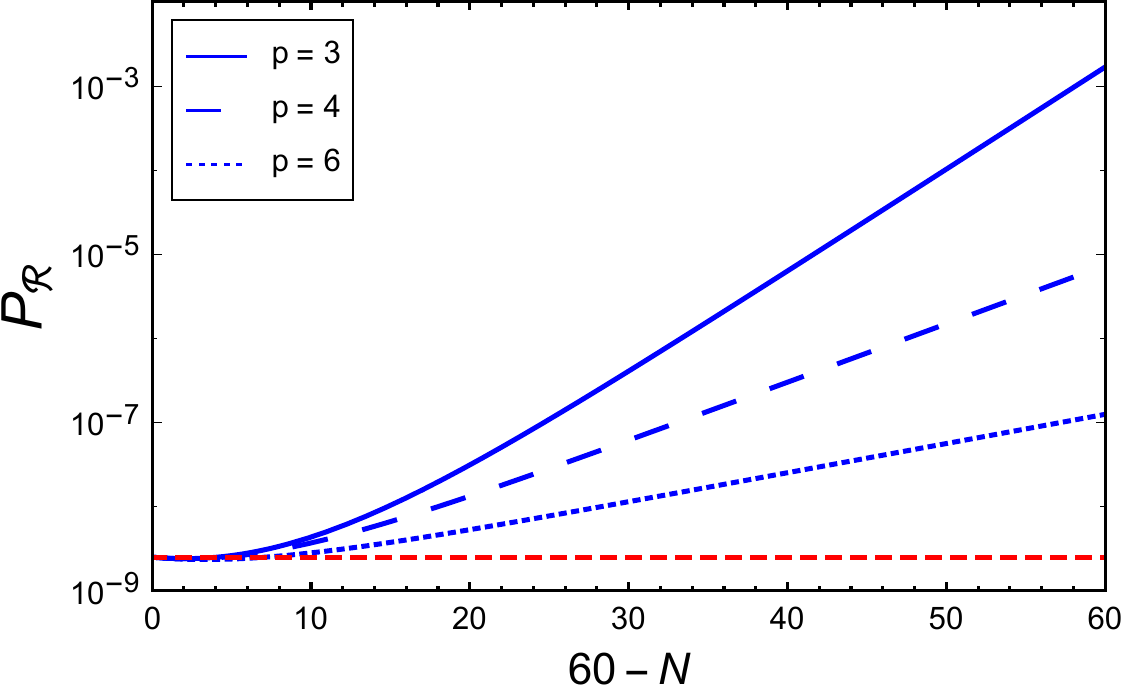}
    \caption{(a) Power spectrum as a function of $\eta_{0}$ at the end of inflation $N=0$ for $p=3,4$ and $6$, respectively. The red dashed line indicates the spectrum amplitude at the CMB scale. (b) Power spectrum as a function of e-fold number $N$ where $\eta_{0}=0.14,0.08$ and $0.05$ corresponding to $p=3,4$ and $6$, respectively.}
    \label{fig1}
\end{figure}

In this section, we discuss the power spectrum at the tree level and identify the model parameters that can produce PBHs. The potential in our model becomes flatter near the end of inflation for larger values of $\eta_{0}$, leading to larger spectrum amplitudes. However, the running spectral index sets an upper bound given by $\alpha<0.02$ on the parameter $\eta_{0}$ as shown in Fig \ref{fig0}. Using Eqs.~(\ref{x}) and (\ref{lambda}), the power spectrum can be expressed as a function of $\eta_{0}$ while keeping $p$ fixed: 
\begin{equation}
 P_{\mathcal{R}}(N) = 2.5\times10^{-9} e^{-2\eta_{0}(N+N_{\rm cmb}(p-2))} \Big[ \frac{e^{\eta_{0}(p-2)N}(0.04+2\eta_{0})+e^{\eta_{0}(p-2)N_{\rm cmb}}(2\eta_{0}(p-2)-0.04)}{2\eta_{0}(p-1)} \Big]^{\frac{2p-2}{p-2}}.
 \label{pe}
\end{equation}
The running spectral index takes the form 
\begin{equation}
    \alpha(N) = \frac{2\eta_{0}^2 (p-1)(p-2)(2\eta_{0}+0.04)(2\eta_{0}(p-2)-0.04)e^{\eta_{0}(p-2)(N+N_{\rm cmb})}}{\Big[ (2\eta_{0}+0.04)e^{\eta_{0}(p-2)N}+ (2\eta_{0}(p-2)-0.04)e^{\eta_{0}(p-2)N_{\rm cmb}}  \Big]^2}.
\end{equation} \newline
 We consider three cases with $p=3$, $p=4$, and $p=6$, plotting $P_{\mathcal{R}}$ and $\alpha$ as functions of $\eta_{0}$ and $N$ in Figs. \ref{fig0} and \ref{fig1}, respectively, with $N_T = N_{\rm cmb} = 60$. Both $P_{\mathcal{R}}$ and $\alpha$ increase monotonically with $\eta_{0}$ and upper bound on $\alpha$ constrains the value of $\eta_{0}$. Thus, the spectrum critically depends on this upper bound.
For the model $p=4$ ($p=6$), this constraint gives $\eta_0 \lesssim 0.08$ ($\eta_0 \lesssim 0.05$) and the maximum spectrum at the end of inflation is $P_{\mathcal{R}} \sim 10^{-6}~(\sim 10^{-7})$, which are insufficient to produce PBHs. However, for the model $p=3$, the upper bound of $\eta_0 \lesssim 0.14$ gives $P_{\mathcal{R}} \sim 0.001$, which are large enough to produce PBHs. Therefore, the model $p = 3$ is a promising candidate for PBH production, as it satisfies CMB constraints while producing sufficiently large amplitude of curvature perturbations. In the next section, we will estimate quantum loop corrections in this case. 
\section{Loop effects from the inflaton field and the waterfall field}
\label{loop1}
The higher-order derivative terms of the potential induce loop corrections through the self-interactions of the inflaton field. The power spectrum in Eq.~(\ref{P_R}) directly probes the potential slow roll parameter $\epsilon_{1v}$. One would expect the loop corrections to modify $\epsilon_{1v}$, which will give the leading order contributions to $P_{\mathcal{R}}$. The inflaton field can be split into the homogeneous mean field $\phi(t)$ and its quantum fluctuations $\delta \phi(\mathbf{x},t)$ as 
\begin{equation}
 \phi(\mathbf{x},t) \rightarrow \phi(t)+\delta \phi(\mathbf{x},t).
\end{equation}
The quantum correction to the equation of motion of the inflaton mean field is given by
\begin{equation}
    \Ddot{\phi}(t) + 3 H(t) \Dot{\phi}(t) + V_{\phi}^{\prime} (\phi) + \frac{V_{\phi}'''(\phi) }{2} \big< \delta \phi(\mathbf{x},t)^2 \big> +\dots = 0. \label{eominqm}
\end{equation} 
 The Fourier modes of $\delta \phi$ satisfy the  Mukhanov-Sasaki equation given by
 \begin{equation}\label{mode_eq}
      \delta \ddot \phi_k + 3H(t)  \delta \dot\phi_k + \frac{k^2}{a^2} \delta \phi_k + \Bigg[ V^{\prime \prime}_{\phi}(\phi) - \frac{1}{a^3} \dfrac{\mathrm d}{\mathrm dt} \Big( \frac{a^3 \dot \phi^2}{H}  \Big) \Bigg] \delta \phi_k = 0.
\end{equation} 
Consequently, this suggests the identification of the effective potential \cite{boyanovsky2006quantum}
\begin{equation} \label{veff}
    V_{\phi, eff}(\phi) \equiv V_{\phi}(\phi) + \frac{V_{\phi}''(\phi)}{2}\big< \delta \phi^2(\mathbf{x},t) \big>.
\end{equation}

Using the effective one-loop corrected equation of motion under slow-roll approximation, we define the effective first slow roll parameter \cite{sloth2006one,sloth2007one}
\begin{equation}
  \epsilon_{1eff}\equiv  \frac{1}{2} \left( \frac{V_{\phi ,eff}'(\phi)}{V_{\phi,eff}(\phi)}\right)^2 \approx \frac{1}{2} \left( \frac{V_{\phi}'}{V_{\phi}}\right)^2 +\frac{1}{2} \left( \frac{V_{\phi}'}{V_{\phi}}\right)^2 \left(\frac{V_{\phi}^{\prime\prime\prime}}{V_{\phi}^{\prime}}-\frac{V_{\phi}^{\prime\prime}}{V_{\phi}} \right)\big< \delta \phi^2(\mathbf{x},t) \big>  \equiv \epsilon_{1v} + \delta\epsilon_{1\phi},
    \end{equation}
 where the correction to the first slow-roll parameter is given by
\begin{align}
    \delta\epsilon_{1\phi} &= \frac{1}{2} \left( \frac{V_{\phi}'}{V_{\phi}}\right)^2 \left(\frac{V_{\phi}^{\prime\prime\prime}}{V_{\phi}^{\prime}}-\frac{V_{\phi}^{\prime\prime}}{V_{\phi}} \right)\big< \delta \phi^2(\mathbf{x},t) \big>.  \label{sr1loop}
\end{align}
 Similarly, we define the effective second slow roll parameter as
\begin{align}
\epsilon_{2eff} &\equiv \frac{V_{\phi, eff}''}{V_{\phi, eff}} \approx  \frac{V_{\phi}''}{V_{\phi}} +  \frac{V_{\phi}''}{V_{\phi}}
\left(\frac{V_{\phi}^{[4]}}{2V_{\phi}^{\prime\prime}}-\frac{V_{\phi}^{\prime\prime}}{2V_{\phi}}\right)\big< \delta \phi^2(\mathbf{x},t) \big> \equiv \epsilon_{2v} + \delta \epsilon_{2\phi},
\end{align}
with the correction to the second slow-roll parameter
\begin{equation}
\delta \epsilon_{2\phi} = \frac{V_{\phi}''}{V_{\phi}}
\left(\frac{V_{\phi}^{[4]}}{2V_{\phi}^{\prime\prime}}-\frac{V_{\phi}^{\prime\prime}}{2V_{\phi}}\right)\big< \delta \phi^2(\mathbf{x},t) \big>.
\end{equation}
The field fluctuations are given by
\begin{equation}\label{deltaphi}
    \big<\delta \phi(\mathbf{x},\eta)^2\big> = \displaystyle\int \frac{d^3k}{(2\pi)^3} \frac{|f_{k}(\eta)|^2}{a^2} = \displaystyle\int_{0}^{\infty}\frac{dk}{k} P_{\mathcal{\delta \phi}}(k,\eta)
\end{equation} where $f_{k}=a \delta \phi_{k}$ and $\eta=-\frac{1}{Ha}$ is the conformal time in de-Sitter spacetime. The power spectrum   \begin{equation}
    P_{\mathcal{\delta \phi}}(k,\eta) =  \frac{k^3}{2\pi^2} \frac{|f_{k}(\eta)|^2}{a^2}=\frac{H^2}{4\pi^2} \left( 1+ (k \eta)^2\right)
\end{equation} under slow roll approximation is calculated using the Bunch-Davies mode function in de-Sitter spacetime
\begin{equation}
    f_{k}(\eta) = \frac{\sqrt{\pi}}{2}\sqrt{-\eta} H^{(1)}_{\frac{3}{2}}(-k\eta) e^{+i{\pi}}.
\end{equation}
Then, the curvature perturbation in (\ref{P_R}) is recovered for the superhorizon modes in the limit of $- k\eta \rightarrow 0$ or $k\ll aH$. Now, the equation (\ref{deltaphi}) becomes
\begin{equation}
     \big<\delta \phi(\mathbf{x},\eta)^2\big> = \displaystyle\int_{0}^{\infty}\frac{dk}{k} \frac{H^2}{4\pi^2} (1+(-k\eta)^2).
\end{equation} This integral contains both UV and IR divergence. The UV part of the divergence can be removed through a renormalization procedure \cite{boyanovsky2006quantum}, and we introduce the IR cutoff relevant to observations to evaluate the integral. The IR part of the spectrum  corresponding to the superhorizon limit $k\ll aH$  at the end of inflation $\eta=\eta_e$ gives 
\begin{equation}
    \big<\delta \phi(\mathbf{x},\eta_e)^2\big>_{IR} \approx \displaystyle\int_{k_{i}}^{k_{e}}\frac{dk}{k} \frac{H^2}{4\pi^2} \approx \frac{V_{0}}{12\pi^2} N_{T}
\end{equation} where the IR cutoff $k_{i}=a_{i}H_{i}$ is given by the physical scale that leaves the horizon at the beginning of inflation \cite{sloth2006one,sloth2007one} and $k_{e}$ corresponds to the physical scale that leaves the horizon at the end of inflation with the Hubble parameter  $H_i \sim H_e=\sqrt{V_0/3}$.
If we assume $N_T=N_{\rm cmb}=60$ e-folds of inflation for the model $p=3$ with $\eta_0=0.14$, the fractional change in the first slow-roll parameter at the end is thus of the order of
\begin{equation}
  \frac{\delta \epsilon_{1\phi}}{\epsilon_{1v}}  \approx \frac{N_T}{12 \pi ^2} \left(6 \lambda  \phi _e-\frac{6 \lambda  V_{0}}{\eta_0  V_{0} \phi _e-3 \lambda  \phi _e^2}-\eta_0 V_{0} \right) \approx -1\times 10^{-6},
\end{equation} and the change to the second slow-roll parameter is of the order
\begin{equation}
    \frac{\delta \epsilon_{2\phi}}{\epsilon_{2v}} \approx
    - \frac{N_T}{24 \pi^2}\left(\eta_{0}  V_{0}-6 \lambda  \phi _e\right) \approx -1.0 \times 10^{-11},
\end{equation}
with the parameters $V_0\sim 10^{-10}$, $\lambda\sim 10^{-11}$ and $\phi_e\sim 10^{-4}$ determined from (\ref{x}) and (\ref{lambda}). Although the amplitude of curvature perturbations is large at the end of inflation due to the smallness of $\epsilon_{1v}$, the amplitude of field fluctuations remains small, and scale-invariant. The quantum loop corrections $\big<\delta \phi(\mathbf{x},\eta_e)^2\big>_{IR}$ at the end of inflation is proportional to the total e-fold number $N_T$. We estimate loop corrections for 60 e-folds of inflation, ensuring the model satisfies CMB constraints while producing a large enough amplitude of curvature perturbations without overproducing PBHs. Additionally, the tree-level predictions are unaffected by the loop corrections of order $10^{-6}.$ This makes it a promising candidate for PBH formation without the need to violate the slow-roll approximation. 

We now consider the loop effects from the coupling between the inflaton field and the waterfall field, given by $g^2\phi^2 \psi^2$ as seen from Eq.~(\ref{hy2}). The loop correction from this interaction modifies the equation of motion to 
\begin{equation} \label{eof_m_psi}
\ddot{\phi}+3H\dot{\phi}+V_\phi^\prime(\phi)+2g^2\phi\langle \delta \psi^2 \rangle+\cdots=0.
\end{equation}
 We identify the effective potential as
\begin{equation}
    V_{eff} \equiv V_{\phi}(\phi) + g^2 \phi^2  \big< \delta \psi^2(\mathbf{x},t)\big>.
\end{equation} Thus, the effective slow roll parameters can be obtained from $ V_{eff}$ as
\begin{equation}
    \epsilon_{1veff} \equiv \epsilon_{1v} +  \delta \epsilon_{1\psi}
\end{equation} with the correction to the first slow roll parameter given by
\begin{equation}
    \delta \epsilon_{1\psi} =  \frac{1}{2} \left( \frac{V_{\phi}'}{V_{\phi}}\right)^2 \left(\frac{4 g^2 \phi}{V_{\phi}^{\prime}}-\frac{ 2g^2 \phi^2}{V_{\phi}} \right)\big< \delta \psi^2(\mathbf{x},t) \big>,
\end{equation} and the correction to the second slow roll parameter
\begin{equation}
    \delta \epsilon_{2\psi} \equiv  \left( \frac{V_{\phi}''}{V_{\phi}}\right) \left(\frac{2 g^2 }{V_{\phi}^{\prime\prime}}-\frac{g^2 \phi^2}{V_{\phi}} \right)\big< \delta \psi^2(\mathbf{x},t) \big>.
\end{equation}
The waterfall field fluctuations during inflation are decoupled from the metric perturbation at linear order. The equation of motion is given by \cite{gordon2000adiabatic}
\begin{align}
     \delta \ddot\psi_{k} &+ 3H  \delta \dot\psi_k + \Bigg[ \frac{k^2}{a^2}  +
 M^2_{\psi}(\phi)  \Bigg]\delta \psi_{k} =0,
\end{align} where $M_{\psi}^2$ is the effective mass of the waterfall field as given in Eq.($\ref{effectivemass}$)
\begin{equation}
    \frac{M_{\psi}^2(\phi)}{H^2}  = \frac{12}{\Lambda^2} \left(\frac{\phi^2}{\phi_{e}^2}-1\right) = \beta \left(\frac{\phi^2}{\phi_{e}^2}-1\right)\equiv \Tilde{\beta},
\end{equation}
with $\beta=12/\Lambda^2$ and we have used $H^2\simeq V_{0}/3$ as the vacuum energy dominates. Rewriting the mode equation in conformal time
and introducing new variable $u_{k} \equiv a \delta \psi_{k}$, we have
\begin{equation}
    u_{k}'' + \Big( k^2  +  \frac{\Tilde{\beta}}{\eta^2} \Big)u_{k} =0.
\end{equation} As $\beta \gg 1$, we can approximate $\Tilde{\beta} \sim \beta$ when $\phi>\phi_{e}$. Since $\phi$ slow-rolls down the potential, this equation changes adiabatically, and one can use WKB approximation with the Bunch-Davies vacuum for initial time $\eta\to -\infty $ to obtain \cite{abolhasani2011no}:
 \begin{equation}
     u_{k}(\eta) \simeq \frac{1}{\sqrt[4]{4k^2+4\frac{\Tilde{\beta}}{\eta^2}}} e^{ \pm i \int \sqrt{k^2+\frac{\Tilde{\beta}}{\eta^2}} d\eta}.
 \end{equation} The expectation value of the waterfall field fluctuations are
 \begin{align}
     \big<\delta\psi^2\big>
     = \displaystyle\int \frac{dk}{k} \frac{k^3}{2\pi^2}\Big|\frac{u_{k}}{a}\Big|^2
    \simeq \displaystyle\int_{0}^{\infty} \frac{dk}{k} \frac{H^2}{4\pi^2} \frac{k^3\eta^3}{\sqrt{\Tilde{\beta}+ k^2\eta^2}}.
 \end{align} The power spectrum of the waterfall field fluctuations evaluated at horizon crossing time is suppressed by a factor of $\Tilde{\beta}^{1/2}$ during inflation ($\phi>\phi_{e}$) \cite{abolhasani2011no}. We then focus on the IR part of the integral evaluated at the end of inflation. We consider the modes that leave the horizon during inflation from $k_i$ to $k_e$
 \begin{equation}
      \big<\delta\psi^2\big>_{IR}(\eta_e)
      \simeq \frac{V_{0}}{12\pi^2}\displaystyle\int_{k_i}^{k_e} \frac{dk}{k}  \frac{k^3\eta_e^3}{\sqrt{\Tilde{\beta}+ k^2\eta_e^2}}
          \approx \frac{V_{0}}{12\pi^2} \frac{1}{3\sqrt{\Tilde{\beta}}} \approx \frac{V_{0}}{12\pi^2} \frac{\Lambda}{3\sqrt{12}}\, ,
      \end{equation}
 where $| k_e \eta_e|=1$ and  $| k_i \eta_e|\propto e^{-N_T} \rightarrow 0$ for $N_T=60$ have been used in doing the integral.
In addition, we approximate $\Tilde{\beta} \sim \beta$ to obtain an order of magnitude estimate. This approximation is valid during slow-roll inflation, as the mass squared term $ M_{\psi}^2(\phi)$ remains positive, ensuring the reliability of the WKB approximation.

 The fractional change in the first slow-roll parameter just before the end of inflation for the model $p=3$ with $\eta_{0}=0.14$ is given by
 \begin{equation}
   \frac{  \delta\epsilon_{1\psi}}{\epsilon_{1v}} 
   \approx \left(\frac{4 g^2 \phi _e}{\eta_{0}  V_{0} \phi _e-3 \lambda  \phi _e^2}-\frac{2 g^2 \phi _e^2}{V_{0}}\right)\ \Big(\frac{V_{0}}{12\pi^2} \frac{\Lambda}{3\sqrt{12}}\Big)\, .
 \end{equation}
 
We have $\lambda \sim 10^{-11}$ and $V_{0} \sim 10^{-10}$. Using equations (\ref{g}), (\ref{xe}) and (\ref{x}), we obtain $\phi_{e} \sim 10^{-4}$, as well as $g \Lambda \sim 10^{-1}$. The coupling constant $g^2<1$ for the vacuum-dominated regime \cite{copeland1994false} and to justify perturbation methods. The loop corrections are proportional to $g$, and to have rapid instability triggered at the end of 60 e-folds of inflation, we choose $g=0.926574$.  Then, the change in the first slow-roll parameter is of order 
 \begin{equation}
   \frac{  \delta\epsilon_{1\psi}}{\epsilon_{1v}}  \sim \mathcal{O}(10^{-3}),
 \end{equation}
and the change in the second slow-roll parameter is
\begin{equation}
    \frac{\delta \epsilon_{2\psi}}{\epsilon_{2v}}  \approx \left( \frac{2 g^2}{\eta_{0}  V_{0}-6 \lambda  \phi _e}-\frac{g^2 \phi _e^2}{V_{0}} \right) \Big( \frac{V_{0}}{12\pi^2} \frac{\Lambda}{3\sqrt{12}}\Big) \sim \mathcal{O}(10^{-3}).
\end{equation} We have estimated the loop corrections for the model $p=3$ from the waterfall field interaction to be of order $10^{-3}$ and from self-interaction of the inflation field to be of order $10^{-6}$.  Therefore, quantum corrections are negligible in both cases and do not affect the tree-level results.

\subsection*{ Primordial Black Hole Mass} 
In this subsection, we follow the analytical model \footnote{We note that the estimation scheme of PBH abundance and hence the mass has developed well beyond the simple method in \cite{Carr:1975qj}(see \cite{escriva2024primordial} and references therein). From the compaction function approach and numerical simulations, it is well-known that the PBH mass features a scaling relation $M \propto \mathcal{K}(\delta - \delta_{c})^{\gamma}$, that depends on the shape of the curvature fluctuation and equation of state (see \cite{escriva2024primordial} and references therein). We neglect these corrections in this work as we do not consider the precise statistics of curvature perturbation beyond the power spectrum and leave accurate estimation to future work. } from \cite{Carr:1975qj} to estimate the PBH mass associated with the peak of the power spectrum. Large curvature perturbations generated during inflation lead to significant density perturbations after inflation. These overdensities collapse into PBHs if they exceed a certain threshold when the corresponding perturbation modes re-enter the horizon. The characteristic mass of the PBHs is related to the total energy enclosed within the Hubble horizon at re-entry:
    \begin{equation}
        M_{PBH} = \gamma M_{H} \Big|_{k=aH} \simeq \Big(\frac{g_{*}}{10.75}\Big)^{-1/6} \Big[ \frac{k}{4.22\times10^{6} ~\text{Mp}c^{-1}} \Big]^{-2} \textup{M}_\odot\  ,
    \end{equation} where $M_{H}(t) =  \frac{4 \pi \rho(t)}{3H(t)^3}$ is the time-dependent mass, $\gamma~(\simeq 0.2)$ is the numerical constant that accounts for the efficiency of collapse and $g_{*}$ is the number of relativistic degrees of freedom in the energy density and assumed to be equal to that of entropy density. We use the entropy conservation and the energy density scaling for the plasma temperature during the radiation-dominated era in the second equality to obtain the horizon mass when the $k$-mode enters the horizon \cite{Nakama:2016gzw,escriva2024primordial, Choudhury:2024aji}.
\begin{figure}[t]
  \centering
\includegraphics[width=0.75\textwidth]{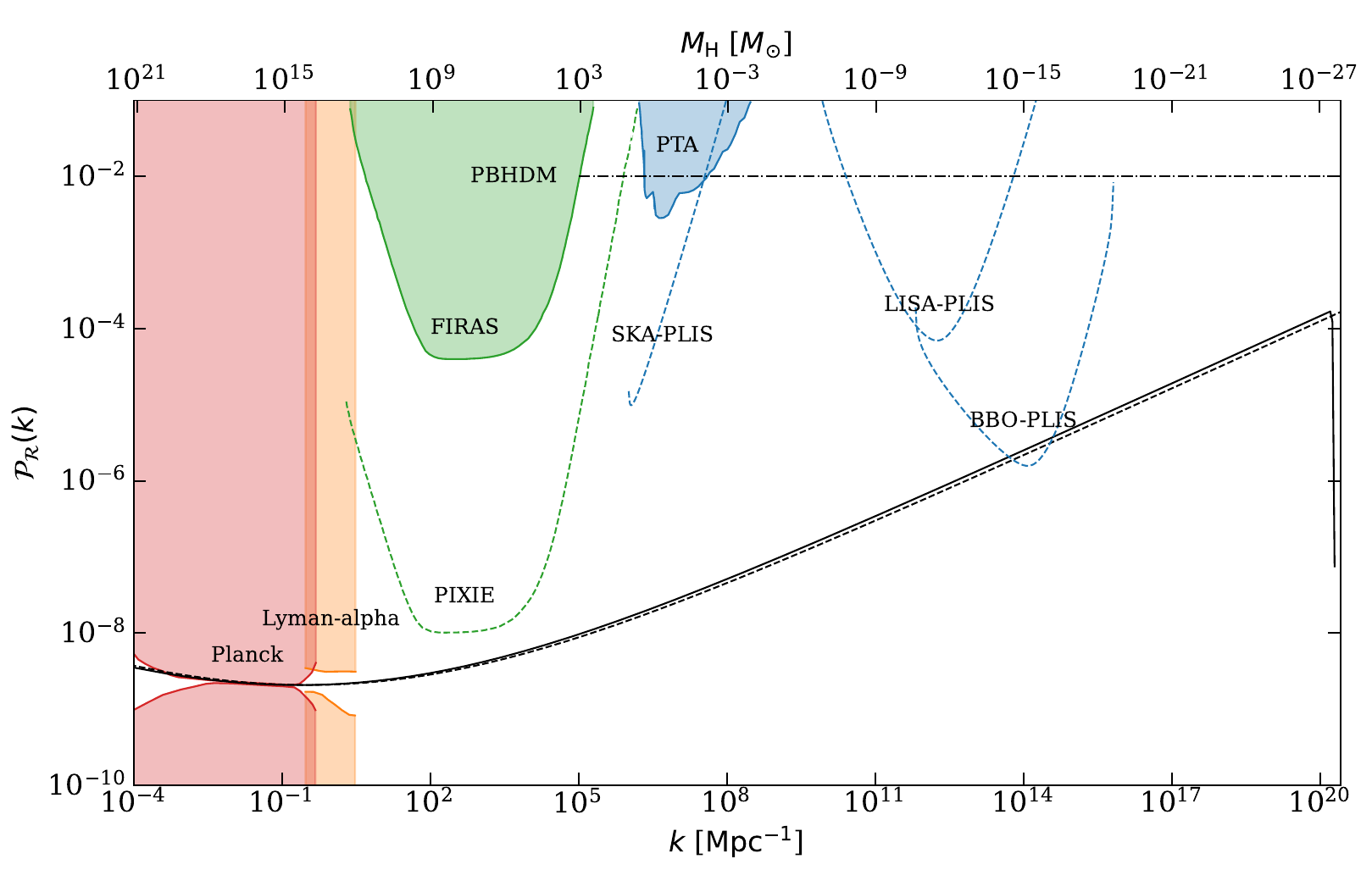}
  \caption{ The power spectrum of curvature perturbation calculated from solving two-field evolution of background equations is represented by a black solid line (the dashed line indicates the spectrum from numerical simulation of single field approximation). The amplitude and spectral index is set at $N=50$ for the case $p=3, \eta_{0}=0.14$ with $(\phi_{i},\psi_{i})=(0.8112,7\times10^{-6})$ for a better fit of Planck data \cite{akrami2020planck}. Constraints on primordial power spectrum from the Plank 2018 data \cite{akrami2020planck} (red), the Lyman$-\alpha$ forest \cite{bird2011minimally} (orange), CMB spectral distortions (FIRAS) \cite{Fixsen:1996nj, Mather:1993ij} (green) and pulsar timing array limits on gravitational waves \cite{Byrnes:2018txb} (blue) are shown where in each case the experimentally excluded regions are shaded. The approximate amplitude of $10^{-2}$ required to form an interesting number of PBHs is shown as a black dash-dot line (PBHDM) and potential constraints from future experiments like PIXIE (green) and limits on gravitational waves from the power-law-integrated sensitivities (PLIS) of SKA, LISA, and BBO (blue) \cite{Inomata:2018epa, Chluba:2019nxa,schmitz_2020_3689582} are shown as dotted lines. The figure is adapted from \cite{bradley_j_kavanagh_2019_3538999}.}
  \label{PBH_bounds}
\end{figure}
The mass associated with the peak at $k=10^{20}~\text{Mpc}^{-1}$ is of order $10^{-27} \textup{M}_\odot\ $ or $10^{6}g$. PBHs of mass less than $10^{15} g$ would have evaporated by now through Hawking radiation. Those in the range $10^{10}-10^{15}g$ can be constrained by Big Bang nucleosynthesis (BBN), CMB anisotropies, the Galactic and extra-galactic $\gamma$-ray and cosmic ray backgrounds \cite{Carr:2020gox}. Since the mass associated with the peak of the power spectrum is much less than $10^{10}g$, the produced PBHs would have evaporated before BBN, leaving no observational signature. \\
However, the spectrum has a relatively broad peak that falls within the power-law-integrated detector sensitivity  \cite{schmitz_2020_3689582} of future experiments like The Big Bang Observer (BBO) over a range of values (see Fig. \ref{PBH_bounds}), indicating the possibility of testing this model by measuring the scalar-induced gravitational waves (SIGW) generated by large curvature perturbations. While this was shown in reference \cite{alabidi2012observable} for a single-field hilltop model, we note that further calculations are required to account for the overall shape of the curvature power spectrum and the role of the waterfall field in our model to obtain detailed quantitative estimates of the expected amplitudes and detectability of the SIGW signals, which we address in our future work. \\
 To produce massive PBHs in this model, one should either break the slow-roll approximation in the rapid instability case or consider a mild waterfall transition. If the waterfall field transition is mild, the isocurvature perturbations can significantly contribute to the amplitude of the power spectrum, potentially leading to the formation of massive PBHs \cite{clesse2015massive, kawasaki2016can, tada2023primordial}. 

\section{Conclusion}
We analytically calculate the power spectrum in Type III hilltop models under the slow-roll approximation following \cite{kohri2007more}. We impose CMB constraints to obtain conditions on model parameters. The spectrum depends on parameters $\eta_0$ and $N$, and for a fixed total number of e-folds, it monotonically increases with $\eta_{0}$. However, the running spectral index sets an upper bound on $\eta_{0}$ and for the model $p=3$ with $\eta_0 \lesssim 0.14$, the spectrum at the end of inflation is as large as $P_{\mathcal{R}} \sim 0.001$. Therefore, the model $p = 3$ is a promising candidate for primordial black hole production, as it satisfies CMB constraints while producing a sufficiently large amplitude of curvature perturbations without overproducing PBHs.

We then consider loop corrections in the model $p=3$ from both the self-interaction of inflation and its interaction with the waterfall field by evaluating the fractional change in the slow-roll parameters. The loop corrections from the self-interaction are proportional to the number of e-folds $N_{T}$, and for 60 e-folds of inflation, corrections are of order $\sim 10^{-6}$. The corrections from the waterfall field interaction for the rapid insatiably case where inflation ends as soon as $M_{\psi}^2(\phi)$ becomes negative, are of order $\sim 10^{-3}$.

The loop corrections from both cases are negligible and do not affect the tree-level results, supporting the model $p = 3$ as a promising candidate for PBH formation within the slow-roll approximation. The mass of the produced PBHs  ($\sim10^{6}g$) corresponding to the peak of the spectrum, in this case, is too small to be observed. However, the spectrum has a relatively broad peak that falls within the detector sensitivity of future experiments like the BBO (see Fig. 5), indicating the possibility of testing this model by measuring the SIGW signals. We note that further calculations are required to provide precise quantitative estimates of the amplitude and detectability of these SIGW signals, which we address in our future work. Our ongoing work extends to the second phase of inflation in the waterfall regime, focusing on the two-field dynamics described in \cite{gordon2000adiabatic} to study quantum one-loop corrections and explore the possibility of producing massive PBHs.

\acknowledgments
This work is supported by the National Science and Technology Council (NSTC) of Taiwan under grant numbers 112-2112-M-167-001-MY2 (CML), 111-2112-M-259 -006 -MY3 (DSL) and 112-2112-M-259-016 (CPY).
\bibliographystyle{unsrtnat}  
\bibliography{references}  

\begin{thebibliography}{55}
\providecommand{\natexlab}[1]{#1}
\providecommand{\url}[1]{\texttt{#1}}
\expandafter\ifx\csname urlstyle\endcsname\relax
  \providecommand{\doi}[1]{doi: #1}\else
  \providecommand{\doi}{doi: \begingroup \urlstyle{rm}\Url}\fi

\bibitem[Liddle and Lyth(2000)]{liddle2000cosmological}
Andrew~R Liddle and David~H Lyth.
\newblock \emph{Cosmological inflation and large-scale structure}.
\newblock Cambridge university press, 2000.

\bibitem[Starobinsky(1980)]{starobinsky1980new}
Alexei~A Starobinsky.
\newblock A new type of isotropic cosmological models without singularity.
\newblock \emph{Physics Letters B}, 91\penalty0 (1):\penalty0 99--102, 1980.

\bibitem[Guth(1981)]{guth1981inflationary}
Alan~H Guth.
\newblock Inflationary universe: A possible solution to the horizon and flatness problems.
\newblock \emph{Physical Review D}, 23\penalty0 (2):\penalty0 347, 1981.

\bibitem[Linde(1982)]{linde1982new}
Andrei~D Linde.
\newblock A new inflationary universe scenario: a possible solution of the horizon, flatness, homogeneity, isotropy and primordial monopole problems.
\newblock \emph{Physics Letters B}, 108\penalty0 (6):\penalty0 389--393, 1982.

\bibitem[Kohri et~al.(2007)Kohri, Lin, and Lyth]{kohri2007more}
Kazunori Kohri, Chia-Min Lin, and David~H Lyth.
\newblock More hilltop inflation models.
\newblock \emph{Journal of Cosmology and Astroparticle Physics}, 2007\penalty0 (12):\penalty0 004, 2007.

\bibitem[Alabidi et~al.(2012)Alabidi, Kohri, Sasaki, and Sendouda]{alabidi2012observable}
Laila Alabidi, Kazunori Kohri, Misao Sasaki, and Yuuiti Sendouda.
\newblock Observable spectra of induced gravitational waves from inflation.
\newblock \emph{Journal of Cosmology and Astroparticle Physics}, 2012\penalty0 (09):\penalty0 017, 2012.

\bibitem[Alabidi and Kohri(2009)]{alabidi2009generating}
Laila Alabidi and Kazunori Kohri.
\newblock Generating primordial black holes via hilltop-type inflation models.
\newblock \emph{Physical Review D—Particles, Fields, Gravitation, and Cosmology}, 80\penalty0 (6):\penalty0 063511, 2009.

\bibitem[Kristiano and Yokoyama(2024)]{kristiano2024constraining}
Jason Kristiano and Jun’ichi Yokoyama.
\newblock Constraining primordial black hole formation from single-field inflation.
\newblock \emph{Physical Review Letters}, 132\penalty0 (22):\penalty0 221003, 2024.

\bibitem[Riotto(2023{\natexlab{a}})]{riotto2023primordial}
Antonio Riotto.
\newblock The primordial black hole formation from single-field inflation is not ruled out.
\newblock \emph{arXiv preprint arXiv:2301.00599}, 2023{\natexlab{a}}.

\bibitem[Riotto(2023{\natexlab{b}})]{riotto2023primordialblackholeformation}
A.~Riotto.
\newblock The primordial black hole formation from single-field inflation is still not ruled out, 2023{\natexlab{b}}.
\newblock URL \url{https://arxiv.org/abs/2303.01727}.

\bibitem[Choudhury et~al.(2023{\natexlab{a}})Choudhury, Gangopadhyay, and Sami]{choudhury2023no}
Sayantan Choudhury, Mayukh~R Gangopadhyay, and M~Sami.
\newblock No-go for the formation of heavy mass primordial black holes in single field inflation.
\newblock \emph{arXiv preprint arXiv:2301.10000}, 2023{\natexlab{a}}.

\bibitem[Choudhury et~al.(2023{\natexlab{b}})Choudhury, Panda, and Sami]{choudhury2023pbh}
Sayantan Choudhury, Sudhakar Panda, and M~Sami.
\newblock Pbh formation in eft of single field inflation with sharp transition.
\newblock \emph{Physics Letters B}, 845:\penalty0 138123, 2023{\natexlab{b}}.

\bibitem[Choudhury et~al.(2023{\natexlab{c}})Choudhury, Panda, and Sami]{choudhury2023quantum}
Sayantan Choudhury, Sudhakar Panda, and M~Sami.
\newblock Quantum loop effects on the power spectrum and constraints on primordial black holes.
\newblock \emph{Journal of Cosmology and Astroparticle Physics}, 2023\penalty0 (11):\penalty0 066, 2023{\natexlab{c}}.

\bibitem[Firouzjahi(2023)]{firouzjahi2023one}
Hassan Firouzjahi.
\newblock One-loop corrections in power spectrum in single field inflation.
\newblock \emph{Journal of Cosmology and Astroparticle Physics}, 2023\penalty0 (10):\penalty0 006, 2023.

\bibitem[Firouzjahi and Riotto(2024)]{firouzjahi2024primordial}
Hassan Firouzjahi and Antonio Riotto.
\newblock Primordial black holes and loops in single-field inflation.
\newblock \emph{Journal of Cosmology and Astroparticle Physics}, 2024\penalty0 (02):\penalty0 021, 2024.

\bibitem[Franciolini et~al.(2023)Franciolini, Iovino, Taoso, and Urbano]{franciolini2023one}
Gabriele Franciolini, Antonio~Junior Iovino, Marco Taoso, and Alfredo Urbano.
\newblock One loop to rule them all: Perturbativity in the presence of ultra slow-roll dynamics.
\newblock \emph{arXiv preprint arXiv:2305.03491}, 2023.

\bibitem[Fumagalli(2023)]{fumagalli2023absence}
Jacopo Fumagalli.
\newblock Absence of one-loop effects on large scales from small scales in non-slow-roll dynamics.
\newblock \emph{arXiv preprint arXiv:2305.19263}, 2023.

\bibitem[Syu et~al.(2020)Syu, Lee, and Ng]{syu2020quantum}
Wei-Can Syu, Da-Shin Lee, and Kin-Wang Ng.
\newblock Quantum loop effects to the power spectrum of primordial perturbations during ultra slow-roll inflation.
\newblock \emph{Physical Review D}, 101\penalty0 (2):\penalty0 025013, 2020.

\bibitem[Cheng et~al.(2022)Cheng, Lee, and Ng]{cheng2022power}
Shu-Lin Cheng, Da-Shin Lee, and Kin-Wang Ng.
\newblock Power spectrum of primordial perturbations during ultra-slow-roll inflation with back reaction effects.
\newblock \emph{Physics Letters B}, 827:\penalty0 136956, 2022.

\bibitem[Cheng et~al.(2024)Cheng, Lee, and Ng]{cheng2024primordial}
Shu-Lin Cheng, Da-Shin Lee, and Kin-Wang Ng.
\newblock Primordial perturbations from ultra-slow-roll single-field inflation with quantum loop effects.
\newblock \emph{Journal of Cosmology and Astroparticle Physics}, 2024\penalty0 (03):\penalty0 008, 2024.

\bibitem[Stewart(1997{\natexlab{a}})]{stewart1997flattening}
Ewan~D Stewart.
\newblock Flattening the inflaton's potential with quantum corrections.
\newblock \emph{Physics Letters B}, 391\penalty0 (1-2):\penalty0 34--38, 1997{\natexlab{a}}.

\bibitem[Stewart(1997{\natexlab{b}})]{Stewart:1997wg}
Ewan~D. Stewart.
\newblock {Flattening the inflaton's potential with quantum corrections. 2.}
\newblock \emph{Phys. Rev. D}, 56:\penalty0 2019--2023, 1997{\natexlab{b}}.
\newblock \doi{10.1103/PhysRevD.56.2019}.

\bibitem[Boyanovsky et~al.(2006)Boyanovsky, de~Vega, and Sanchez]{boyanovsky2006quantum}
D~Boyanovsky, Hector~J de~Vega, and Norma~G Sanchez.
\newblock Quantum corrections to slow roll inflation and new scaling of superhorizon fluctuations.
\newblock \emph{Nuclear Physics B}, 747\penalty0 (1-2):\penalty0 25--54, 2006.

\bibitem[Sloth(2006)]{sloth2006one}
Martin~S Sloth.
\newblock On the one loop corrections to inflation and the cmb anisotropies.
\newblock \emph{Nuclear Physics B}, 748\penalty0 (1-2):\penalty0 149--169, 2006.

\bibitem[Sloth(2007)]{sloth2007one}
Martin~S Sloth.
\newblock On the one-loop corrections to inflation ii: The consistency relation.
\newblock \emph{Nuclear Physics B}, 775\penalty0 (1-2):\penalty0 78--94, 2007.

\bibitem[Linde(1994)]{linde1994hybrid}
Andrei Linde.
\newblock Hybrid inflation.
\newblock \emph{Physical Review D}, 49\penalty0 (2):\penalty0 748, 1994.

\bibitem[Copeland et~al.(1994)Copeland, Liddle, Lyth, Stewart, and Wands]{copeland1994false}
Edmund~J Copeland, Andrew~R Liddle, David~H Lyth, Ewan~D Stewart, and David Wands.
\newblock False vacuum inflation with einstein gravity.
\newblock \emph{Physical Review D}, 49\penalty0 (12):\penalty0 6410, 1994.

\bibitem[Clesse and Garc{\'\i}a-Bellido(2015)]{clesse2015massive}
S{\'e}bastien Clesse and Juan Garc{\'\i}a-Bellido.
\newblock Massive primordial black holes from hybrid inflation as dark matter and the seeds of galaxies.
\newblock \emph{Physical Review D}, 92\penalty0 (2):\penalty0 023524, 2015.

\bibitem[Kawasaki and Tada(2016)]{kawasaki2016can}
Masahiro Kawasaki and Yuichiro Tada.
\newblock Can massive primordial black holes be produced in mild waterfall hybrid inflation?
\newblock \emph{Journal of Cosmology and Astroparticle Physics}, 2016\penalty0 (08):\penalty0 041, 2016.

\bibitem[Tada and Yamada(2023)]{tada2023primordial}
Yuichiro Tada and Masaki Yamada.
\newblock Primordial black hole formation in hybrid inflation.
\newblock \emph{Physical Review D}, 107\penalty0 (12):\penalty0 123539, 2023.

\bibitem[Clesse(2011)]{clesse2011hybrid}
Sebastien Clesse.
\newblock Hybrid inflation along waterfall trajectories.
\newblock \emph{Physical Review D—Particles, Fields, Gravitation, and Cosmology}, 83\penalty0 (6):\penalty0 063518, 2011.

\bibitem[Kodama et~al.(2011)Kodama, Kohri, and Nakayama]{Kodama:2011vs}
Hideo Kodama, Kazunori Kohri, and Kazunori Nakayama.
\newblock {On the waterfall behavior in hybrid inflation}.
\newblock \emph{Prog. Theor. Phys.}, 126:\penalty0 331--350, 2011.
\newblock \doi{10.1143/PTP.126.331}.

\bibitem[Akrami et~al.(2020)Akrami, Arroja, Ashdown, Aumont, Baccigalupi, Ballardini, Banday, Barreiro, Bartolo, Basak, et~al.]{akrami2020planck}
Yashar Akrami, Frederico Arroja, M~Ashdown, J~Aumont, Carlo Baccigalupi, M~Ballardini, Anthony~J Banday, RB~Barreiro, Nicola Bartolo, S~Basak, et~al.
\newblock Planck 2018 results-x. constraints on inflation.
\newblock \emph{Astronomy \& Astrophysics}, 641:\penalty0 A10, 2020.

\bibitem[Lin(2024)]{lin2024hilltop}
Chia-Min Lin.
\newblock Hilltop sneutrino hybrid inflation.
\newblock \emph{Chinese Journal of Physics}, 87:\penalty0 138--144, 2024.

\bibitem[Kohri et~al.(2014)Kohri, Lim, Lin, and Mimura]{kohri2014hilltop}
Kazunori Kohri, CS~Lim, Chia-Min Lin, and Yukihiro Mimura.
\newblock Hilltop supernatural inflation and susy unified models.
\newblock \emph{Journal of Cosmology and Astroparticle Physics}, 2014\penalty0 (01):\penalty0 029, 2014.

\bibitem[Lin and Cheung(2009)]{lin2009reducing}
Chia-Min Lin and Kingman Cheung.
\newblock Reducing the spectral index in supernatural inflation.
\newblock \emph{Physical Review D—Particles, Fields, Gravitation, and Cosmology}, 79\penalty0 (8):\penalty0 083509, 2009.

\bibitem[Lin(2021)]{lin2021testing}
Chia-Min Lin.
\newblock Testing hilltop supernatural inflation with gravitational waves.
\newblock \emph{Journal of Cosmology and Astroparticle Physics}, 2021\penalty0 (05):\penalty0 056, 2021.

\bibitem[Lin and Kohri(2020)]{lin2020inflaton}
Chia-Min Lin and Kazunori Kohri.
\newblock Inflaton as the affleck-dine baryogenesis field in hilltop supernatural inflation.
\newblock \emph{Physical Review D}, 102\penalty0 (4):\penalty0 043511, 2020.

\bibitem[Kohri and Lin(2010)]{kohri2010hilltop}
Kazunori Kohri and Chia-Min Lin.
\newblock Hilltop supernatural inflation and gravitino problem.
\newblock \emph{Journal of Cosmology and Astroparticle Physics}, 2010\penalty0 (11):\penalty0 010, 2010.

\bibitem[Gordon et~al.(2000)Gordon, Wands, Bassett, and Maartens]{gordon2000adiabatic}
Christopher Gordon, David Wands, Bruce~A Bassett, and Roy Maartens.
\newblock Adiabatic and entropy perturbations from inflation.
\newblock \emph{Physical Review D}, 63\penalty0 (2):\penalty0 023506, 2000.

\bibitem[Wolf(2024)]{PhysRevD.110.043521}
William~J. Wolf.
\newblock Minimizing the tensor-to-scalar ratio in single-field inflation models.
\newblock \emph{Phys. Rev. D}, 110:\penalty0 043521, Aug 2024.
\newblock \doi{10.1103/PhysRevD.110.043521}.
\newblock URL \url{https://link.aps.org/doi/10.1103/PhysRevD.110.043521}.

\bibitem[Abolhasani and Firouzjahi(2011)]{abolhasani2011no}
Ali~Akbar Abolhasani and Hassan Firouzjahi.
\newblock No large scale curvature perturbations during the waterfall phase transition of hybrid inflation.
\newblock \emph{Physical Review D—Particles, Fields, Gravitation, and Cosmology}, 83\penalty0 (6):\penalty0 063513, 2011.

\bibitem[Carr(1975)]{Carr:1975qj}
Bernard~J. Carr.
\newblock {The Primordial black hole mass spectrum}.
\newblock \emph{Astrophys. J.}, 201:\penalty0 1--19, 1975.
\newblock \doi{10.1086/153853}.

\bibitem[Escriv{\`a} et~al.(2024)Escriv{\`a}, K{\"u}hnel, and Tada]{escriva2024primordial}
Albert Escriv{\`a}, Florian K{\"u}hnel, and Yuichiro Tada.
\newblock Primordial black holes.
\newblock In \emph{Black Holes in the Era of Gravitational-Wave Astronomy}, pages 261--377. Elsevier, 2024.

\bibitem[Nakama et~al.(2017)Nakama, Silk, and Kamionkowski]{Nakama:2016gzw}
Tomohiro Nakama, Joseph Silk, and Marc Kamionkowski.
\newblock {Stochastic gravitational waves associated with the formation of primordial black holes}.
\newblock \emph{Phys. Rev. D}, 95\penalty0 (4):\penalty0 043511, 2017.
\newblock \doi{10.1103/PhysRevD.95.043511}.

\bibitem[Choudhury and Sami(2024)]{Choudhury:2024aji}
Sayantan Choudhury and M.~Sami.
\newblock {Large fluctuations and Primordial Black Holes}, 7 2024.

\bibitem[Bird et~al.(2011)Bird, Peiris, Viel, and Verde]{bird2011minimally}
Simeon Bird, Hiranya~V Peiris, Matteo Viel, and Licia Verde.
\newblock Minimally parametric power spectrum reconstruction from the lyman $\alpha$ forest.
\newblock \emph{Monthly Notices of the Royal Astronomical Society}, 413\penalty0 (3):\penalty0 1717--1728, 2011.

\bibitem[Fixsen et~al.(1996)Fixsen, Cheng, Gales, Mather, Shafer, and Wright]{Fixsen:1996nj}
D.~J. Fixsen, E.~S. Cheng, J.~M. Gales, John~C. Mather, R.~A. Shafer, and E.~L. Wright.
\newblock {The Cosmic Microwave Background spectrum from the full COBE FIRAS data set}.
\newblock \emph{Astrophys. J.}, 473:\penalty0 576, 1996.
\newblock \doi{10.1086/178173}.

\bibitem[Mather et~al.(1994)]{Mather:1993ij}
John~C. Mather et~al.
\newblock {Measurement of the Cosmic Microwave Background spectrum by the COBE FIRAS instrument}.
\newblock \emph{Astrophys. J.}, 420:\penalty0 439--444, 1994.
\newblock \doi{10.1086/173574}.

\bibitem[Byrnes et~al.(2019)Byrnes, Cole, and Patil]{Byrnes:2018txb}
Christian~T. Byrnes, Philippa~S. Cole, and Subodh~P. Patil.
\newblock {Steepest growth of the power spectrum and primordial black holes}.
\newblock \emph{JCAP}, 06:\penalty0 028, 2019.
\newblock \doi{10.1088/1475-7516/2019/06/028}.

\bibitem[Inomata and Nakama(2019)]{Inomata:2018epa}
Keisuke Inomata and Tomohiro Nakama.
\newblock {Gravitational waves induced by scalar perturbations as probes of the small-scale primordial spectrum}.
\newblock \emph{Phys. Rev. D}, 99\penalty0 (4):\penalty0 043511, 2019.
\newblock \doi{10.1103/PhysRevD.99.043511}.

\bibitem[Chluba et~al.(2021)]{Chluba:2019nxa}
J.~Chluba et~al.
\newblock {New horizons in cosmology with spectral distortions of the cosmic microwave background}.
\newblock \emph{Exper. Astron.}, 51\penalty0 (3):\penalty0 1515--1554, 2021.
\newblock \doi{10.1007/s10686-021-09729-5}.

\bibitem[Schmitz(2020)]{schmitz_2020_3689582}
Kai Schmitz.
\newblock New sensitivity curves for gravitational-wave experiments, February 2020.
\newblock URL \url{https://doi.org/10.5281/zenodo.3689582}.

\bibitem[Kavanagh(2019)]{bradley_j_kavanagh_2019_3538999}
Bradley~J. Kavanagh.
\newblock bradkav/pbhbounds: Release version, November 2019.
\newblock URL \url{https://doi.org/10.5281/zenodo.3538999}.

\bibitem[Carr et~al.(2021)Carr, Kohri, Sendouda, and Yokoyama]{Carr:2020gox}
Bernard Carr, Kazunori Kohri, Yuuiti Sendouda, and Jun'ichi Yokoyama.
\newblock {Constraints on primordial black holes}.
\newblock \emph{Rept. Prog. Phys.}, 84\penalty0 (11):\penalty0 116902, 2021.
\newblock \doi{10.1088/1361-6633/ac1e31}.

\end{thebibliography}

\end{document}